\documentclass[%
reprint,
nofootinbib,
amsmath,amssymb,
aps,
]{revtex4-2}

\usepackage{graphicx}
\usepackage{dcolumn}
\usepackage{bm}
\usepackage{multirow}
\renewcommand\Re{\operatorname{Re}}
\usepackage{booktabs}
\usepackage{enumitem}

\usepackage{mathptmx}
\usepackage{textcomp}
\usepackage[normalem]{ulem} 

\usepackage{color}

\begin{document}


\title{Characterization of the low electric field and zero-temperature two-level-system loss in hydrogenated amorphous silicon}

\author{Fabien Defrance}
\email[]{fabien.m.defrance@jpl.nasa.gov}
\affiliation{Jet Propulsion Laboratory, \\ California Institute of Technology, Pasadena, CA, USA, 91109}

\author{Andrew D. Beyer}
\affiliation{Jet Propulsion Laboratory, \\ California Institute of Technology, Pasadena, CA, USA, 91109}

\author{Shibo Shu}
\affiliation{California Institute of Technology, Pasadena, CA, USA, 91125}

\author{Jack Sayers}
\affiliation{California Institute of Technology, Pasadena, CA, USA, 91125}

\author{Sunil R. Golwala}
\affiliation{California Institute of Technology, Pasadena, CA, USA, 91125}

\date{December 1, 2023}

\begin{abstract}
Two-level systems (TLS) are an important, if not dominant, source of loss and noise for superconducting resonators such as those used in kinetic inductance detectors and some quantum information science platforms.  They are similarly important for loss in photolithographically fabricated superconducting mm-wave/THz transmission lines.  For both lumped-element and transmission-line structures, native amorphous surface oxide films are typically the sites of such TLS in non-microstripline geometries, while loss in the (usually amorphous) dielectric film itself usually dominates in microstriplines.  We report here on the demonstration of low TLS loss at GHz frequencies in hydrogenated amorphous silicon (a-Si:H) films deposited by plasma-enhanced chemical vapor deposition in superconducting lumped-element resonators using parallel-plate capacitors (PPCs).  The values we obtain from two recipes in different deposition machines, 7$\,\times\,10^{-6}$ and 12$\,\times\,10^{-6}$, improve on the best achieved in the literature by a factor of 2--4 for a-Si:H and are comparable to recent measurements of amorphous germanium.  
Moreover, we have taken care to extract the true  zero-temperature, low-field loss tangent of these films, accounting for temperature and field saturation effects that can yield misleading results.  Such robustly fabricated and characterized films render the use of PPCs with deposited amorphous films a viable architecture for superconducting resonators, and they also promise extremely low loss and high quality factor for photolithographically fabricated superconducting mm-wave/THz transmission lines used in planar antennas and resonant filters.

\vspace{6mm}

\copyright  2023. All rights reserved.

\end{abstract}

\maketitle


\section{Introduction}
\label{sec:intro}

The standard tunneling model (STM) is the reference model used to describe the properties of amorphous dielectrics at low temperatures (below a few Kelvins)~\cite{Phillips:1987, Esquinazi:1998}.  According to this model, amorphous dielectrics have defect states with two physical configurations (\textit{e.g.}, locations of an atom or groups of atoms) with different energies, forming a quantum two-state (``two-level'') system with inter-state tunneling~\cite{Muller:2019, Phillips:1987}.  When the two-level systems (TLSs) present in amorphous dielectrics have an electric dipole moment, they can couple to an oscillating electric field present in the dielectric (in superconductive resonators used in qubits or kinetic inductance detectors, for example) and convert some of the energy stored in the electric field into phonon emission, resulting in dielectric loss and noise.  
In superconducting qubits, dielectric loss is a major source of decoherence, as first inferred by Martinis et al.~\cite{Martinis:2005}.  
In kinetic inductance detectors, dielectric loss adds noise and can degrade responsivity and multiplexability~\cite{Zmuidzinas:2012,Siegel:2015}, especially in parallel-plate capacitor (PPC) geometries that are desirable in some applications.  In superconducting transmission line used in photolithographically fabricated planar antennas and resonant bandpass filters at mm-wave/THz frequencies, dielectric loss determines attenuation length~\cite{Chang:2015} and limits spectral resolution~\cite{Shirokoff:2014}. 
Therefore, low-loss dielectrics are highly desirable for diverse low-temperature applications.  

Crystalline dielectrics have very low dielectric loss because of their
low density of defects~\cite{Weber:2011}, but incorporation of crystalline dielectrics into elements like PPCs or superconducting microstripline transmission lines requires complex fabrication techniques~\cite{Denis:2009, Beyer:2017}.  
When simpler techniques are used to fabricate the single-layer equivalents (transmission line resonators using coplanar waveguide (CPW), lumped element resonators using interdigitated capacitors (IDCs), superconducting CPW transmission line), they exhibit TLS loss due to the formation of oxide at the exposed surfaces of the superconducting material and substrate~\cite{Gao:2008b, Wang:2009, Sage:2011}.  
Choosing superconducting materials having a weak reactivity with oxygen, and thus producing a very thin oxide layer, yields interesting results with materials such as rhenium or nitrides (TiN, NbN, etc.)~\cite{Wang:2009, Sage:2011, Boussaha:2020}.  However, this solution limits designs to use of a few superconductive materials.

By contrast, multi-layer structures incorporating an easily deposited, low-loss, amorphous dielectric would have the benefit of confining the electric field inside the dielectric, drastically limiting the participation of surface oxides that enhance the loss of CPW and IDC structures on crystalline dielectric.
Such multi-layer structures would enable significant progress in a wide range of low-temperature applications.  
KID development would benefit from the option of replacing large IDCs limited by poorly controllable surface-oxide loss with order-of-magnitude smaller PPCs limited by well-controlled bulk dielectric film loss.  Microstripline KID designs would become less challenging to implement (cf.\ \cite{Mirzaei:2020}).
Low-loss microstripline would also enable more sophisticated planar antennas, higher resolution filters in superconductive spectrometers, and microstripline-based traveling-wave parametric amplifiers using kinetic inductance.  (The films discussed in this paper have already been successful for the last application~\cite{Shu:2021}.) 

The main amorphous dielectrics whose GHz--THz \textit{electromagnetic} behavior --- believed to be due to electric-dipole-coupled TLS --- has been explored to date in the literature are AlO\textsubscript{x}, SiO\textsubscript{2}, SiO\textsubscript{x}, SiN\textsubscript{x}, and hydrogenated amorphous silicon (a-Si:H) (see~\cite{McRae:2020RSI} for an exhaustive review).  
Very recently, hydrogenated amorphous silicon carbide (a-SiC:H;~\cite{Buijtendorp:2022}) and amorphous germanium (a-Ge\;~\cite{Kopas:2021}) have been 
explored.\footnote{We neglect epitaxially grown films (\textit{e.g.}, Al\textsubscript{2}O\textsubscript{3}~\cite{Weides:2011, Cho:2013}, Si~\cite{Kopas:2021}) because epitaxy requirements on the substrate or the base film present significant constraints on use in PPCs or microstripline.  }
Prior studies~\cite{OConnell:2008, Mazin:2010, Bruno:2012, Buijtendorp:2020, Hahnle:2021, Buijtendorp:2022} show that a-Si:H dielectric loss can, in general, be at least an order of magnitude lower than that of AlO\textsubscript{x}, SiO\textsubscript{2}, SiO\textsubscript{x},  and SiN\textsubscript{x} (though low loss has been achieved for SiN\textsubscript{x}~\cite{Paik:2010}, albeit with high stress).  Loss in a-Ge films are the lowest seen in the literature to date~\cite{Kopas:2021}.
\textit{In this paper, we demonstrate a-Si:H films with low-power loss tangent 7 and $\mathit{12\,\times\,10^{-6}}$.  These films improve on the best, previous, robust measurements for a-Si:H~\cite{OConnell:2008} by a factor of 2--4~\footnote{The films presented in~\cite{Buijtendorp:2020, Buijtendorp:2022} could, after corrections and with measurements at lower power, yield comparable low-power loss tangent, but precise numbers at low power are not available.  The most robust numbers available for the same deposition process are in~\cite{Hahnle:2021}, $3.6\,\times\,10^{-5}$ (see Section~\ref{sec:comparisons}).}, and are comparable to measurements for a-Ge given their uncertainties~\cite{Kopas:2021}.}

There is significant, potentially relevant literature on a-Si and a-Si:H in three different contexts: as a photovoltaic, as an example of sub-electronic-bandgap optical absorption, with impact on its use as a mirror coating in laser-interferometric gravitational wave detectors; and, as an exemplar of universal STM behavior.  We review this literature in Appendix~\ref{app:review}, leaving it out of the body of the paper because it is not likely to be relevant for reasons explained there, with the exception of the results of Molina-Ruiz~et~al.~\cite{Molina:2021}.  We do offer a caveat to this conclusion in \S\ref{sec:conclusion}.

Demonstrations of low loss have not usually addressed robustness: stability of the deposition recipe over time, transferability between deposition machines, and suitability of the recipe for combination with other fabrication steps.  For example, films made using the \cite{OConnell:2008, Mazin:2010} recipe \textit{in the same machine} a decade later yielded much higher loss tangent, and low-loss SiN\textsubscript{x} suffers high stress~\cite{Paik:2010}, rendering it challenging to incorporate in complex circuits.  \textit{We demonstrate here $\mathit{\approx 10^{-5}}$ low-power loss tangent using two different combinations of deposition technique and machine, and our recipes are, by construction, consistent with inclusion in PPCs and microstripline~\cite{Shu:2021}.}

Another challenge with measurements of such low loss is systematic uncertainties.  
As described in Section~\ref{tls_loss_theory}, the loss tangent at 0~K and weak electric field, $\tan \delta_{TLS}^0$, characterizes the intrinsic TLS loss of the film, with none of the two-level systems saturated by temperature or power.  
Most previous studies of a-Si:H loss tangent characterized the resonator quality factor $Q_i$ and inferred from it the loss tangent via $\tan \delta = Q_i^{-1}$.  The quantity $\tan \delta$ can be smaller than $\tan \delta_{TLS}^0$ because of the above saturation effects.  
It can also be larger because it may include additional non-TLS losses ($\tan \delta_{other}$).  
Moreover, for a-Si:H loss tangent studies using CPW resonators, separating the TLS loss of a-Si:H from that of surface oxide layers can add systematic uncertainty to the measurements of $\tan \delta_{TLS}^0$ (\textit{e.g.}, \cite{McRae:2020RSI, McRae:2020APL}).  
\textit{We present in this article measurements of $tan\ \delta_{TLS}^0$ that avoid these inaccuracies by design and by measurement technique~\footnote{The utility of PPCs has been known for some time~\cite{Martinis:2005} and was emphasized in \cite{OConnell:2008} and, more recently, in~\cite{McRae:2020APL}.  The applicability of the frequency-shift technique was first pointed out in~\cite{Gao:2008, Pappas:2011, Sage:2011}, with \cite{Gao:2008, Pappas:2011} both: 1)~explaining that readout power saturation of TLS does not impact frequency shift vs.\ temperature because it is sensitive to the full ensemble of TLS while quality factor is only sensitive to TLS within a linewidth of $f_{res}$ (temperature saturation is of course the means by which the frequency shift data yield the low temperature loss tangent); and, 2)~recognizing the convenience of the frequency shift technique relative to measuring quality factor at low power.  \cite{McRae:2020RSI} reiterated these points.  This technique is, however, only usable when the superconducting film has high enough $T_c$ that the dependence of kinetic inductance on temperature does not dominate over the TLS effect, a caveat that may explain the prevalence of $Q_i$ measurements in the literature.  In particular, half of the literature results cited in Section~\ref{sec:comparisons} use Al superconductor, which can suffer this effect.}}: our LC resonator device design uses PPCs to ensure surface oxide contributions are negligible; and, our measurement technique fits for the dependence of resonant frequency $f_{res}$ on temperature and of inverse quality factor $Q_i^{-1}$ on readout power, two independent techniques that yield broadly consistent results, with the former being more robust.  

\section{Test Device Design and Fabrication}

\begin{figure*}[t]
    \includegraphics[width=13cm]{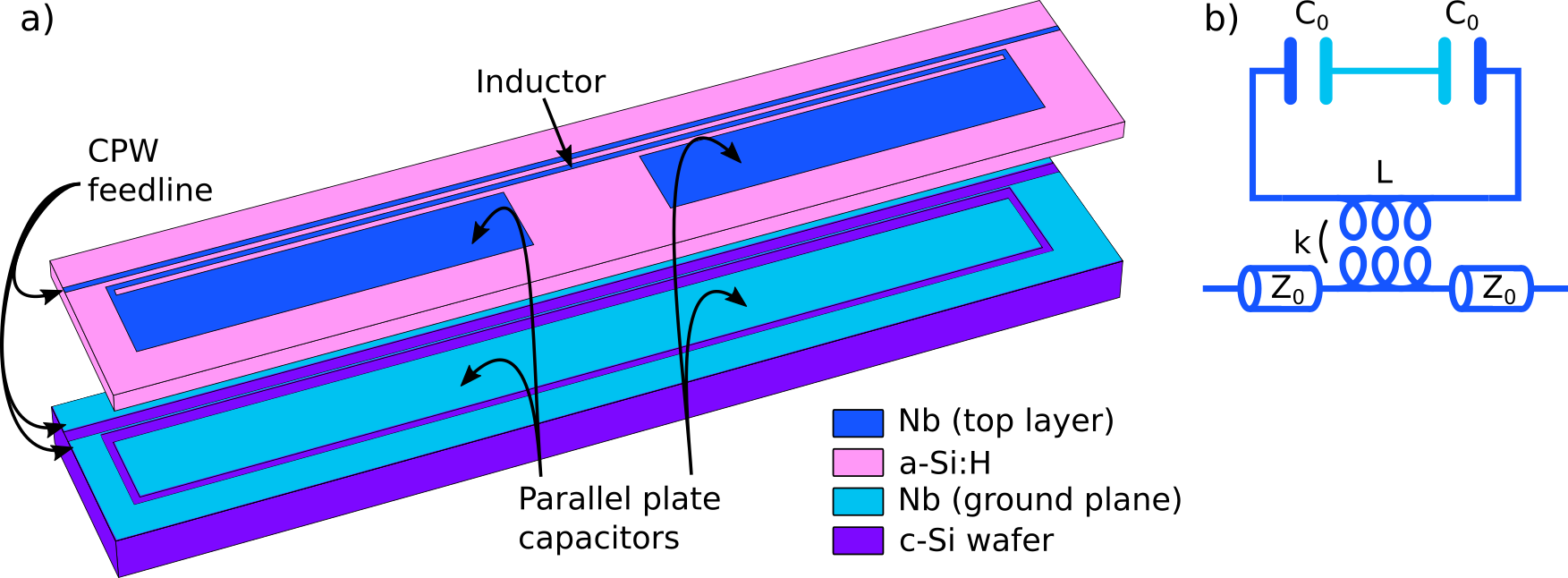}
    \caption{\label{fig:LCschematic} a) Exploded view, not to scale, showing the geometry of the LC resonators used to measure the loss tangent and internal quality factor of a-Si:H.  The CPW feedline is composed of a center conductor on the top Nb layer and ground electrodes on the bottom Nb (``ground plane'') layer separated by the a-Si:H dielectric film to be characterized.  The resonator consists of an inductor that runs parallel to the feedline center conductor on the top Nb layer and a series pair of PPCs formed by plates on the Nb top layer and the ground plane.
    (The a-Si:H layer is only 800~nm thick while the inductor-top plate gap is 22~$\mu$m and the CPW center conductor and gap are 20~$\mu$m and 12~$\mu$m wide, respectively, so the structure is approximately planar in spite of the vertical separation of the center conductor and ground plane.)
    A 44~$\mu$m gap separates the ground plane plate from the surrounding ground plane, though this gap is not strictly necessary because this electrode acts as a virtual ground.  
    The inductor couples the CPW feedline to the LC circuit, and a gap in the ground plane below the inductor mitigates magnetic screening that would reduce the inductance value and feedline coupling.  The gap also limits parasitic capacitance with the ground plane.
    b)~Lumped element circuit equivalent.  The CPW feedline has an impedance $Z_0 = 50\,\Omega$, the two capacitances $C_0$ correspond to the two PPCs, $L$ is the inductance, and $k$ represents the mutual inductance with the CPW.}
\end{figure*}

\subsection{Design}

Each test device hosts 6 niobium (Nb) LC resonators inductively coupled to a 50 $\Omega$ Nb coplanar waveguide (CPW) feedline that is used for readout.  
Each resonator is composed of two parallel plate capacitors (PPCs) in series with an inductor, as described in Figure~\ref{fig:LCschematic}.  
To facilitate their identification, the six resonators are grouped in frequency into two triplets, centered on 0.85~GHz and 1.55~GHz.  
Within each triplet, the resonators are designed to have resonant frequencies separated by 5\%.
The physical dimensions of the resonator components and their predicted/simulated electrical values are listed in Table~\ref{tab:reso_dims}.  Because we initially had a conservative expectation of obtaining TLS loss tangent $\tan \delta \approx 10^{-4}$ and thus internal quality factor $Q_i \approx 10^4$, we designed the coupling quality factor, $Q_c$, to also be close to $10^{4}$ to maximize the coupling efficiency~\cite{Zmuidzinas:2012}.  
$Q_c$ and $f_{res}$ were designed and simulated using Sonnet software and the very useful method developed by Wisbey et al.~\cite{Wisbey:2014}.  
The unexpectedly high values of measured $Q_i$ led to $Q_i \gg Q_c$, causing very deep, overcoupled resonances.  

\begin{table*}[t]
\begin{ruledtabular}
\begin{tabular}{lrrrrrrrr}
     & \multicolumn{3}{c}{PPCs} & \multicolumn{3}{c}{Inductor} & 
     \multicolumn{1}{c}{$f_{res}$} & \multicolumn{1}{c}{$Q_c$} \\ \midrule
     & \multicolumn{1}{c}{$w$} 
     & \multicolumn{1}{c}{$\ell$} 
     & \multicolumn{1}{c}{$C_0$} 
     & \multicolumn{1}{c}{$w$} 
     & \multicolumn{1}{c}{$\ell$} 
     & \multicolumn{1}{c}{$L$} &  &  \\ 
     & \multicolumn{1}{c}{[$\mu$m]} 
     & \multicolumn{1}{c}{[$\mu$m]} 
     & \multicolumn{1}{c}{[pF]} 
     & \multicolumn{1}{c}{[$\mu$m]} 
     & \multicolumn{1}{c}{[$\mu$m]} 
     & \multicolumn{1}{c}{[nH]} 
     & \multicolumn{1}{c}{[MHz]} &  \\ \midrule
    Lower & 265 & 1236 & 42 & 20 & 2920 & 1.84 & 810 & $8 \times 10^3$ \\ 
    frequency & 245 & 1223 & 39 & 20 & 2920 & 1.84 & 844 & $8 \times 10^3$ \\ 
    triplet & 225 & 1210 & 35 & 20 & 2920 & 1.84 & 882 & $8 \times 10^3$ \\ 
    \addlinespace[+1.5ex]
    Higher & 265 & 1236 & 42 & 20 & 760 & 0.54 & 1489 & $2 \times 10^4$ \\ 
    frequency & 245 & 1223 & 39 & 20 & 760 & 0.54 & 1554 & $2 \times 10^4$ \\ 
    triplet & 225 & 1210 & 35 & 20 & 760 & 0.54 & 1627 & $2 \times 10^4$ \\ 
    \end{tabular}
    \end{ruledtabular}
    \caption{\label{tab:reso_dims} Designed dimensions and electrical parameters of the six resonators present on each device.  For the PPCs, $w$ and $\ell$ correspond respectively to the width and length of the top plates, while, for the inductor, $w$ and $\ell$ are the width and length of the inductor line.  
    $C_0$ is the capacitance of each of the two PPCs, so the total capacitance is $C = C_0/2$ ($C_0$ is calculated using the capacitor dimensions and the permitivity of silicon).  The distance between the inductor and the CPW feedline is set to 22~\textmu m to obtain $Q_c \sim 10^4$.}
\end{table*}

\subsection{Fabrication}

\begin{figure}[t]
    \includegraphics[width=8.5cm]{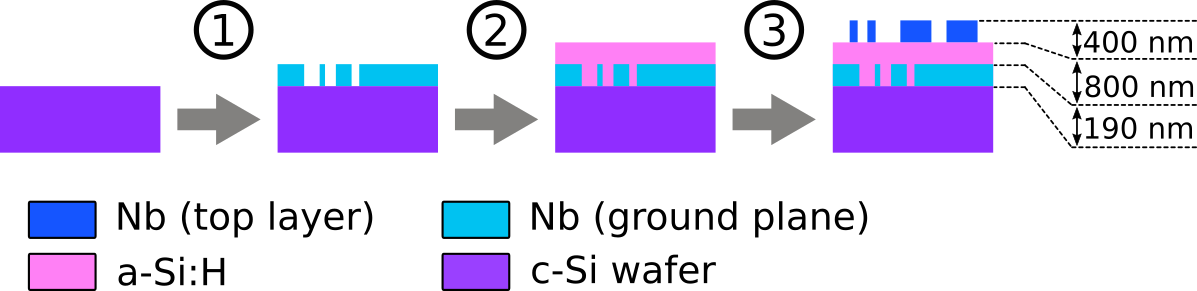}
    \caption{\label{fig:fab} Schematic showing main device fabrication steps.  1)~Deposition and patterning (etch-back) of a 190~nm thick Nb film on a 375~\textmu m thick, 100 mm diameter, high resistivity silicon wafer, forming the ground plane (also including the return conductors of the CPW feedline) and the bottom plate of the PPCs.  2)~Deposition of a 800~nm thick layer of a-Si:H using Plasma-Enhanced Chemical Vapor Deposition (PECVD; Caltech KNI) for recipe A and Inductively Coupled Plasma PECVD (ICP-PECVD; JPL MDL) for recipe B.  3)~Deposition and patterning (etch-back) of a 400~nm thick Nb film on top of the a-Si:H layer, forming the CPW feedline center conductor, the inductor, and the PPC top plates.  We deposit all the metal films using RF magnetron sputtering at JPL MDL with a 6-inch target.}
\end{figure}

\begin{table*}[t]
\begin{ruledtabular}
\begin{tabular}{lccccccccc}
         Recipe & Gas ratio & Temperature & Gas pressure & Gas flow & RF Power & ICP Power & Deposition time & Process & Facility \\ 
         & (SiH\textsubscript{4}/Ar) & {[}\textdegree C{]} & {[}mTorr{]} & {[}sccm{]} & {[}W{]} & {[}W{]} &  &  \\ \midrule
        A & 5\% / 95\% & 350 & 800 & 250 & 10 & N/A & 27'11" & PECVD & KNI \\
        B & 100\% / 0\% & 350 & 10 & 30 & 50 & 300 & 26'06" & ICP-PECVD & JPL \\ 
        \end{tabular}
\end{ruledtabular}
        \caption{\label{tab:fab_recipes} Deposition recipes for a-Si:H.  The gas ratio (SiH\textsubscript{4}/Ar) corresponds to the relative flow rate of the 2 gases.  The two recipes use two different machines and deposition processes.  Recipe A uses an Oxford Plasmalab System 100 at Caltech KNI for PECVD while Recipe B uses an Oxford Plasmalab System 100 ICP 380 at JPL MDL for ICP-PECVD.  Safety restrictions prevented the use of a pure SiH$_4$ atmosphere at Caltech KNI.}
\end{table*}

We fabricated the devices at the NASA Jet Propulsion Laboratory's MicroDevices Laboratory (MDL) and at the Caltech Kavli Nanoscience Institute (KNI) clean room facility.  
For each recipe, we fabricated four devices simultaneously on the same 4-inch high resistivity silicon wafer.  In the rest of this article, we will identify each of the four devices (for each recipe) with an index going from 1 to 4.  Figure~\ref{fig:fab} describes the main fabrication steps.

Previous studies demonstrated that TLS impacting resonator behavior reside primarily in oxide layers localized at interfaces (metal-vacuum, metal-dielectric, and dielectric-vacuum) \cite{Woods:2019, Gao:2008b, wenner:2011, Wang:2009, Sage:2011}.  
In our PPC geometry, the electric field created by the resonator is confined between the capacitor plates and only the metal-dielectric interfaces are relevant.   
Therefore, we took particular care to eliminate any oxide layers at these interfaces: we used buffered oxide etch (BOE) on all silicon surfaces prior to metal-film deposition, we preceded depositions of a-Si:H in the JPL ICP-PECVD machine by Ar\textsuperscript{+} ion milling (not possible in the KNI PECVD machine), and we did the same with depositions of Nb on a-Si:H.  

We tested many different recipes for a-Si:H, mainly varying the gas composition, gas flow, RF power, temperature, and deposition technique.  The two recipes (A \& B; Table~\ref{tab:fab_recipes}) we present here are among those giving the lowest TLS loss values.  
A more detailed review of all the recipes we have tested, with in-depth analysis of a-Si:H structure and composition in addition to measurements of loss tangent, will be published separately.

\section{Theoretical Model for the Effect of Two-Level Systems on Resonator Parameters}
\label{tls_loss_theory}

TLS present in dielectrics can couple to a time-varying electric field via their electric
dipole moments.  They can also emit phonons.  The transfer of energy from the electric field to phonon emission is a form of loss for the dielectric.  
The circuit's total loss can be expressed as \cite{Wang:2015}:
\begin{equation}
    \label{eq:fill_fact1}
    \tan \delta = \tan \delta_{other} + \sum_j F_j  \tan \delta_j.
\end{equation}
with $\tan \delta$ the total loss tangent, $\tan \delta_j$ the loss tangent of each lossy dielectric, $F_j$ the filling factor of each lossy dielectric (indicating the portion of the device's total energy stored in each dielectric material), and $\tan \delta_{other}$ representing additional non-TLS loss mechanisms.  The circuit loss tangent, $\tan \delta$, is equivalent to $Q_i^{-1}$, the inverse of the internal quality factor, which is a quantity easily measurable when the electric circuit is a resonator.  In the rest of this article, all loss tangents are small compared to unity, so we use the approximation $\tan \delta \approx \delta$.  

The dielectric loss tangent $\delta_j$ is not a single number, however, because saturation effects cause it to depend on electric field strength (\textit{i.e.}, stored energy in the resonator, which is determined by readout power) and temperature.  Moreover, TLS affect not just the imaginary part of the dielectric constant --- \textit{i.e.}, cause loss --- but they also affect the real part --- \textit{i.e.}, cause a frequency shift.  At microwave frequencies and low temperatures, the standard tunneling model (STM) yields the following expressions for these two effects (\textit{e.g.},~\cite{Gao:2008}):
\begin{equation}
    \label{eq:loss_tan}
    Q_i^{-1} = \delta_{other} + \sum_j F_j\, \delta_{j,TLS}^0\, \dfrac{\tanh \left(\dfrac{h f_{res}}{2 k_B T}\right)}{\left[1 + \left(\dfrac{|\vec{E_j}|}{E_{c,j}}\right)^2\right]^{\beta_j}},
\end{equation}
\begin{multline}
    \label{eq:gao}
    \frac{f_{res}(T) - f_{res}(0)}{f_{res}(0)} = \\
    \sum_j \frac{F_j\,\delta_{j,TLS}^0}{\pi} \left[\Re \left[ \Psi \left(\frac{1}{2} - \frac{h f_{res}(0)}{2 j \pi k_B T} \right) \right]  - \ln{\left( \frac{h f_{res}(0)}{2 \pi k_B T}\right)} \right],
\end{multline}
with $\delta_{TLS}^0$ the ``intrinsic'' or ``asymptotic'' (zero temperature, low electric field) loss tangent,
$h$ and $k_B$ the Planck and Boltzmann constants respectively, $T$ the temperature of the dielectric, $\vec{E_j}$ the electric field inside the $j$th dielectric film under consideration, $E_{c,j}$ the critical electric field for TLS saturation for the $j$th dielectric film, $\Psi$ the complex digamma function, and $\beta_j$ an exponent determined by the TLS density of states in dielectric film $j$.  
This exponent is 0.5 for a logarithmically uniform density of states \cite{Phillips:1987}, but the experimental literature yields values between 0.15 and 0.35 \cite{Molina:2021, Burnett:2016}.  

Variants on Equation~\ref{eq:gao} are seen in the literature.  
The sign of $h f_{res}(0) / (2j\pi k_B T)$ can be either $-$ or $+$ because $\overline{\Psi(z)} = \Psi(\overline{z})$, and therefore $\Re [\Psi(\overline{z})] = \Re [\Psi(z)]$. The denominator of the last fraction is sometimes written as $k_B T$ only.  
The absence of $2\pi$ arises from the underlying TLS theory \cite{Phillips:1987}, but it results in a frequency offset.  A careful demonstration of this formula by J.~Gao (\cite{Gao:2008}~Appendix G) shows that the denominator of the last fraction must be $2 \pi k_B T$ in order for the right part of the equation to be equal to zero when $T = 0$~K.

Since TLS couple via electric fields, only capacitive elements contribute to TLS loss~\cite{vissers:2012, McRae:2020APL}.  
Because we use PPCs, the electric field is confined between the capacitor plates.  We assume no oxides, and therefore no TLS, remain at the metal-substrate interfaces given the aforementioned oxide-removal steps.  
Though~\cite{McRae:2020APL} argued for the importance, in PPC measurements, of accounting for TLS residing in the parasitic capacitance of the inductor, we argue our geometry differs substantially from the one they present, leading to a much smaller participation of surface oxides.  Specifically, considering the half of the resonator that is at positive voltage during a particular half period: 1)~the vast majority of field lines emanating from the inductor (present because it is not a virtual ground along its entire length) terminate on the ground plane or the PPC bottom plate, passing through the same a-Si:H being tested with the PPC, because the adjacent PPC top plate sits at a larger voltage; 2)~field lines emanating from the PPC top plate strongly prefer to terminate on the PPC bottom plate rather than on the inductor because the former is much closer (800~nm vs.\ 22~$\mu$m); and, 3)~any field lines that do travel from the PPC top plate to the inductor primarily pass through the higher $\epsilon$ a-Si:H rather than through surface oxides.  The same arguments can be made for the half of the resonator that is at negative voltage.  Together, these points imply that: the bulk of the inductor's parasitic capacitance is subject to the same TLS as the PPC; and, the fraction of field lines terminating on the inductor and passing through surface oxides is a small fraction of the already fractionally small parasitic capacitance of the inductor.

We thus assume the entirety of the resonator stored energy resides inside the a-Si:H film, which gives a filling factor $F = 1$ and allows us to rewrite Equations~\ref{eq:loss_tan} and \ref{eq:gao} as:
\begin{equation}
    \label{eq:loss_tan2}
    Q_i^{-1} = \delta_{other} + \delta_{TLS}^0\, \dfrac{\tanh \left(\dfrac{h f_{res}}{2 k_B T}\right)}{\left[1 + \left(\dfrac{|\vec{E}|}{E_{c}}\right)^2\right]^{\beta}},
\end{equation}
\begin{multline}
    \label{eq:gao2}
    \frac{f_{res}(T) - f_{res}(0)}{f_{res}(0)} = \\
    \frac{\delta_{TLS}^0}{\pi} \left[\Re \left[ \Psi \left(\frac{1}{2} - \frac{h f_{res}(0)}{2 j \pi k_B T} \right) \right] - \ln{\left( \frac{h f_{res}(0)}{2 \pi k_B T}\right)} \right].
\end{multline}
Equation~\ref{eq:loss_tan2} implies that, due to their low energy, TLS saturate at high temperature ($h f_{res} \lesssim 2 k_B T$) and under high electric field ($|\vec{E}| \gtrsim E_c$), reducing the dielectric loss \cite{Pappas:2011}.  
A measurement of $Q_i^{-1}$ as a function of temperature and/or electric field can, in principle, be fit to Equation~\ref{eq:loss_tan2} to determine $\delta_{other}$, $E_c$, $\delta_{TLS}^0$, and $\beta$.  
Equation~\ref{eq:gao2} indicates that the frequency shift suffers no field dependence and has a characteristic temperature dependence --- behaviors of great practical utility for robustly determining $\delta_{TLS}^0$, as we will show.

\begin{figure}[t]
    \includegraphics[width=8.5cm]{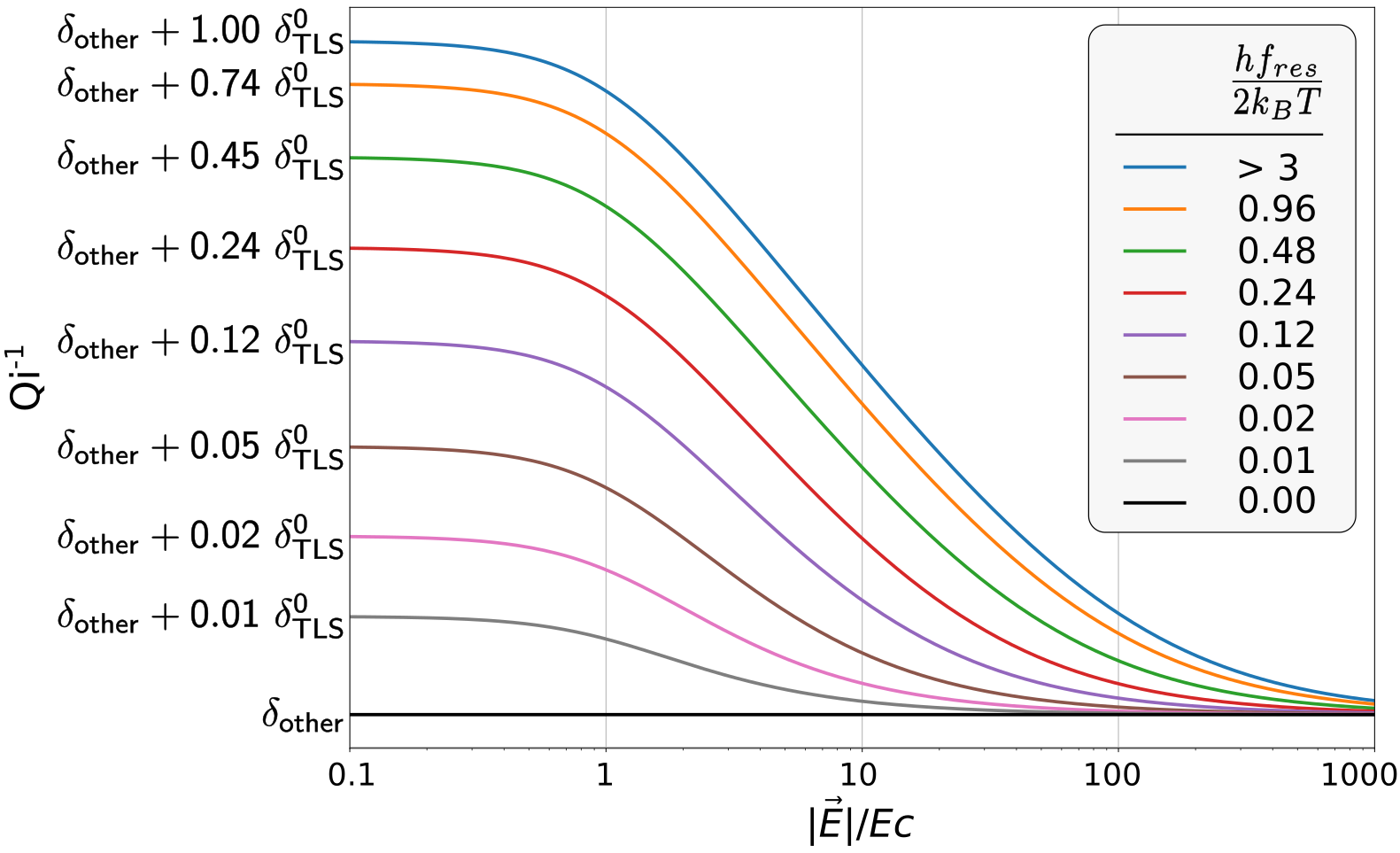}
    \caption{\label{fig:Qi-1_func} Log-log plot illustrating Equation~\ref{eq:loss_tan2}.  $Q_i^{-1}$ converges towards $\delta_{other} + \delta_{TLS}^0\, \tanh (h f_{res}/(2 k_B T))$ at low $|\vec{E}|/E_c$ and towards $\delta_{other}$ at high $|\vec{E}|/E_c$.  The exponent $\beta$ only changes the steepness of the dependence on $|\vec{E}|$; $\beta = 0.5$ was used here, corresponding to a log-uniform TLS density of states.  The different curves correspond to different values of $h f_{res}/(2 k_B T)$.  The plot illustrates that taking data in the fully unsaturated (in $T$ and $|\vec{E}|$) limit yields the best lever arm on determining the true value of $\delta_{TLS}^0$ from $Q_i^{-1}$ data and that taking $T$-saturated data necessitates applying a correction factor $\tanh(h f_{res} / (2 k_B T))$.  
    For reference, $h f_{res} / (2 k_B T) \approx 0.96$ (and $\tanh(0.96) = 0.74)$ for $f_{res} = 1$~GHz at $T = 25$~mK.  Our data probe the range $h f_{res} / (2 k_B T) \in [0.08, 0.16]$, requiring a factor of roughly 6--13 correction for $T$ saturation.
    }
\end{figure}

While fitting to Equation~\ref{eq:loss_tan2} may seem the most straightforward approach to finding $\delta_{TLS}^0$, it can be challenging.  Doing so requires determining $Q_i$, which can be difficult if the complex $S_{21}(f-f_{res})$ trajectory deviates significantly from ideal behavior and/or if $Q_i \gg Q_r$ (the total quality factor $Q_r$ of a resonator is related to its internal quality factor $Q_i$ and its coupling quality factor $Q_c$ by the relation $Q_r^{-1} = Q_i^{-1} + Q_c^{-1}$).  
More importantly, such a fit requires extensive and sensitive data: $|\vec{E}|$ must be swept from $|\vec{E}| \ll E_c$ to $|\vec{E}| \gg E_c$ to fit the $|\vec{E}|$-dependence, these data must show clear plateaus at high and low $|\vec{E}|$ to determine $\delta_{other}$ and the normalization of the $|\vec{E}|$-dependence, and either the data must be taken at $T \ll h\,f_{res}/k_B$ so the TLS are unsaturated or one must apply a correction for the $\tanh (h f_{res} / (2 k_B T))$ dependence.  
Data over a wide range in $T$ would maximize the robustness of the  temperature correction, but taking such data is experimentally time-consuming and not usually done in the literature.  
It may also be difficult to obtain sufficient signal-to-noise at the low readout power required for $|\vec{E}|\ll E_c$, and it may not be possible to reach the $|\vec{E}|\gg E_c$ regime before the superconductor becomes nonlinear (when the current approaches the critical current) or pair-breaking takes place due to dissipation in the superconductor arising from $\delta_{other}$.

In this article, we take the approach --- which have not seen in the literature --- of fitting both Equations~\ref{eq:loss_tan2} and \ref{eq:gao2} and comparing the results in order to better understand the systematics associated with the two methods.  We conclude that the frequency-shift approach (Equation~\ref{eq:gao2}) is far more robust.

\section{Experimental setup}

We cooled the devices using a Chase Cryogenics closed-cycle $^3$He/$^3$He/$^4$He sorption cooler mounted to the second stage of a Cryomech PT415 cryocooler.  Each device resided in its own box sealed to prevent optical radiation from breaking quasiparticles or heating the substrate.  A magnetic shield, residing at 4~K and consisting of two layers of Amuneal A4K material, enclosed the devices to limit the impact of Earth's magnetic field.  A combination of stainless steel and NbTi semi-rigid coaxial cables carried the readout signal  to the devices, with 30~dB and 10~dB in-line attenuators at 4~K and 0.35~K, respectively, to block 300~K thermal noise.  Similar NbTi coax carried the signal exiting each device to a cryogenic low-noise amplifier (LNA) at 4~K, with a noise temperature of approximately 5~K, followed by stainless steel coax back to 300~K.  Additional LNAs at 300~K ensured the cryogenic LNA dominated the system noise.  We monitored the device temperature using a Stanford Research System (SRS) SIM921 reading a Lakeshore Germanium Resistance Thermometer (GRT) located next to the devices.  The temperature was varied between 240~mK and 450~mK using a SRS SIM960 analog PID controller supplying a current to a 10~k$\Omega$ heater on the mechanical stage holding the devices.  A Copper Mountain Technologies SC5065 Vector Network Analyzer (VNA) performed measurements of $S_{21}(f)$ using the above signal chain.  We used the Python module SCRAPS~\cite{Carter:2017} to fit the $S_{21}(f)$ data to standard forms (\textit{e.g.}, \cite{Zmuidzinas:2012}) to extract the resonance frequency $f_{res}$ and quality factors $Q_r$, $Q_i$, and $Q_c$ for each value of $T$ and readout power (electric field).

We varied the readout power applied by the VNA to the feedline, $P_{read}$, over the approximate range $-150$~dBm to $-70$~dBm.  Because the SNR decreases with readout power, we had to compensate by reducing the IF bandwidth of the VNA, which results in a proportional increase in measurement time.  Each factor of 10 decrease in IF bandwidth yielded a 10~dB reduction of noise floor and a factor of 10 increase in measurement time.  We used an IF bandwidth of 1~kHz above $-100$~dBm, reducing it to 10~Hz at the lowest readout powers.

\section{Results}

\begin{figure*}[t]
    \includegraphics[width=17.5cm]{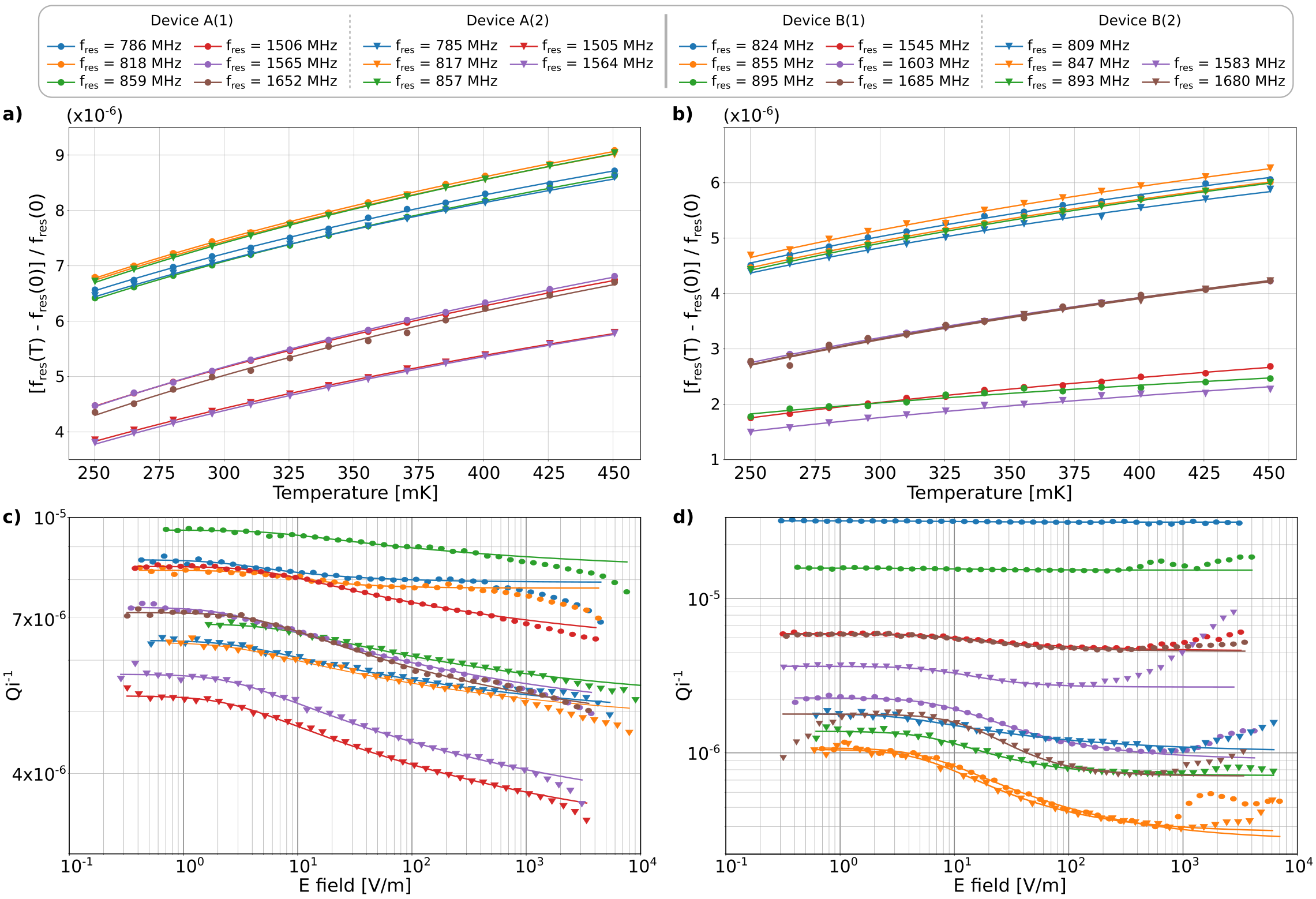}
        \caption{\label{fig:loss_fit} a),~b) Data and best-fit TLS model (Equation~\ref{eq:gao2}) for $(f_{res} - f_{res}(0))/f_{res}(0)$ as a function of temperature.  c),~d)~Data and best-fit TLS model (Equation~\ref{eq:loss_tan2}) for $Q_i^{-1}$ data as a function of electric field.  a),~c)~Devices A(1) and A(2); b),~d)~Devices B(1) and B(2).  Much of the literature uses photon number instead of $|\vec{E}|$; Figure~\ref{fig:loss_fit_vs_Nphotons} in 
        Appendix~\ref{app:photon_number} shows plots c) and d) as a function of photon number to facilitate comparison with published results.  Table~\ref{tab:loss_tan} lists the best-fit parameters.  
        Disagreements between the model and the data are discussed in the text, in particular how we extract $\delta^0_{TLS}$ and $\delta_{other}$ in the presence of such disagreements.}        
        The $f_{res}(T)$ data were taken at a readout power (at the device) of approximately -100~dBm ($|\vec{E}| \approx 500$~V/m).  The $Q_i^{-1}(|\vec{E}|)$ data were taken at 246~mK.
\end{figure*}

\begin{table*}[t]
\begin{ruledtabular}
        \begin{tabular}{@{}r@{\hskip 3mm}r@{\hskip 3mm}r@{\hskip 3mm}r@{\hskip 3mm}r@{\hskip 3mm}r@{\hskip 8mm}r@{\hskip 3mm}r@{\hskip 3mm}r@{\hskip 3mm}r@{\hskip 3mm}r@{\hskip 3mm}r@{}}
        Dataset & \multicolumn{1}{c}{$f_{res}(T)$} & \multicolumn{4}{c}{$Q_i^{-1}(|\vec{E}|)$} & & \multicolumn{1}{c}{$f_{res}(T)$} & \multicolumn{4}{c}{$Q_i^{-1}(|\vec{E}|)$} \\ \midrule
        \multicolumn{1}{c}{$f_{res}$} & \multicolumn{1}{c}{$\delta^0_{TLS}$} & \multicolumn{1}{c}{$\delta^0_{TLS}$} & \multicolumn{1}{c}{$\delta_{other}$} & \multicolumn{1}{c}{$E_c$} & \multicolumn{1}{c}{$\beta$}\hspace{6mm} & \multicolumn{1}{c}{$f_{res}$} & \multicolumn{1}{c}{$\delta^0_{TLS}$} & \multicolumn{1}{c}{$\delta^0_{TLS}$} & \multicolumn{1}{c}{$\delta_{other}$} & \multicolumn{1}{c}{$E_c$} & \multicolumn{1}{c}{$\beta$} \\ \addlinespace[0.5ex]
        \multicolumn{1}{c}{{[}MHz{]}} & ($\times 10^{-6}$) & ($\times 10^{-6}$) & ($\times 10^{-6}$) & \multicolumn{1}{c}{{[}V/m{]}} & & \multicolumn{1}{c}{{[}MHz{]}} & ($\times 10^{-6}$) & ($\times 10^{-6}$) & ($\times 10^{-6}$) & \multicolumn{1}{c}{{[}V/m{]}} & \\ \midrule
        \addlinespace[1ex]
        \multicolumn{6}{l}{Device A(1)} & \multicolumn{6}{l}{Device A(2)} \\ \midrule
        786 & 12.0 & 8.7 & 7.9 & 2.5 & 0.30 & 785 & 11.0 & 21.5 & *4.8 & *2.8 & *0.10 \\
        818 & 12.0 & 6.4 & 7.8 & 6.4 & 0.49 & 817 & 12.0 & 21.0 & *4.7 & *2.6 & *0.10 \\
        859 & 12.0 & 16.2 & *8.2 & *5.7 & *0.10 & 857 & 12.0 & 21.0 & *5.1 & *6.1 & *0.10 \\
        1506 & 12.0 & 15.3 & *6.2 & *4.8 & *0.10 & 1505 & 11.0 & 15.1 & *3.1 & *2.5 & *0.10 \\
        1565 & 13.0 & 16.3 & *4.8 & *2.4 & *0.10 & 1564 & 11.0 & 15.7 & *3.3 & *2.6 & *0.10 \\
        1652 & 13.0 & 16.1 & *4.5 & *4.4 & *0.11 &     &      &      &     &      &      \\
        \bottomrule 
        \addlinespace[1ex]
        \multicolumn{6}{l}{Device B(1)} & \multicolumn{6}{l}{Device B(2)} \\ \midrule
        824 & 8.3 & *7.1 & 28.0 & *4.2 & *0.46 & 809 & 7.9 & 8.9 & 0.9 & 3.4 & 0.20 \\
        855 & 8.3 & 8.7 & 0.2  & 5.8 & 0.27 & 847 & 8.6 & 8.2 & 0.3 & 5.4 & 0.33 \\
        895 & 3.5 & *4.8 & 13.7 & *5.3 & *0.35 & 893 & 8.4 & 6.9 & 0.6 & 7.3 & 0.40 \\
        1545 & 5.0 & 8.1 & 4.1 & 7.2 & 0.27 &      &     &     &     &      &      \\
        1603 & 8.1 & 7.9 & 0.8 & 8.3 & 0.32 & 1583 & 4.4 & 5.7 & 2.4 & 9.4 & 0.60 \\
        1685 & 8.3 & 7.3 & 4.1 & 8.2 & 0.35 & 1680 & 8.3 & 6.0 & 0.6 & 15.4 & 0.60 \\
\end{tabular}
\end{ruledtabular}
          \caption{Best-fit parameters for fits to data shown in Figure~\ref{fig:loss_fit}.  Two resonances are missing in Devices A(2) and B(2) due to fabrication yield.  The values with an asterisk ($^*$) are considered unreliable.  The text explains the fits in more detail and assesses the reliability of the fitted parameters.  Of particular interest is the consistency of $\delta_{TLS}^0$ obtained from the frequency-shift and quality-factor data.
    }
          \label{tab:loss_tan}
\end{table*}

\begin{figure}[t]
    \includegraphics[width=8.5cm]{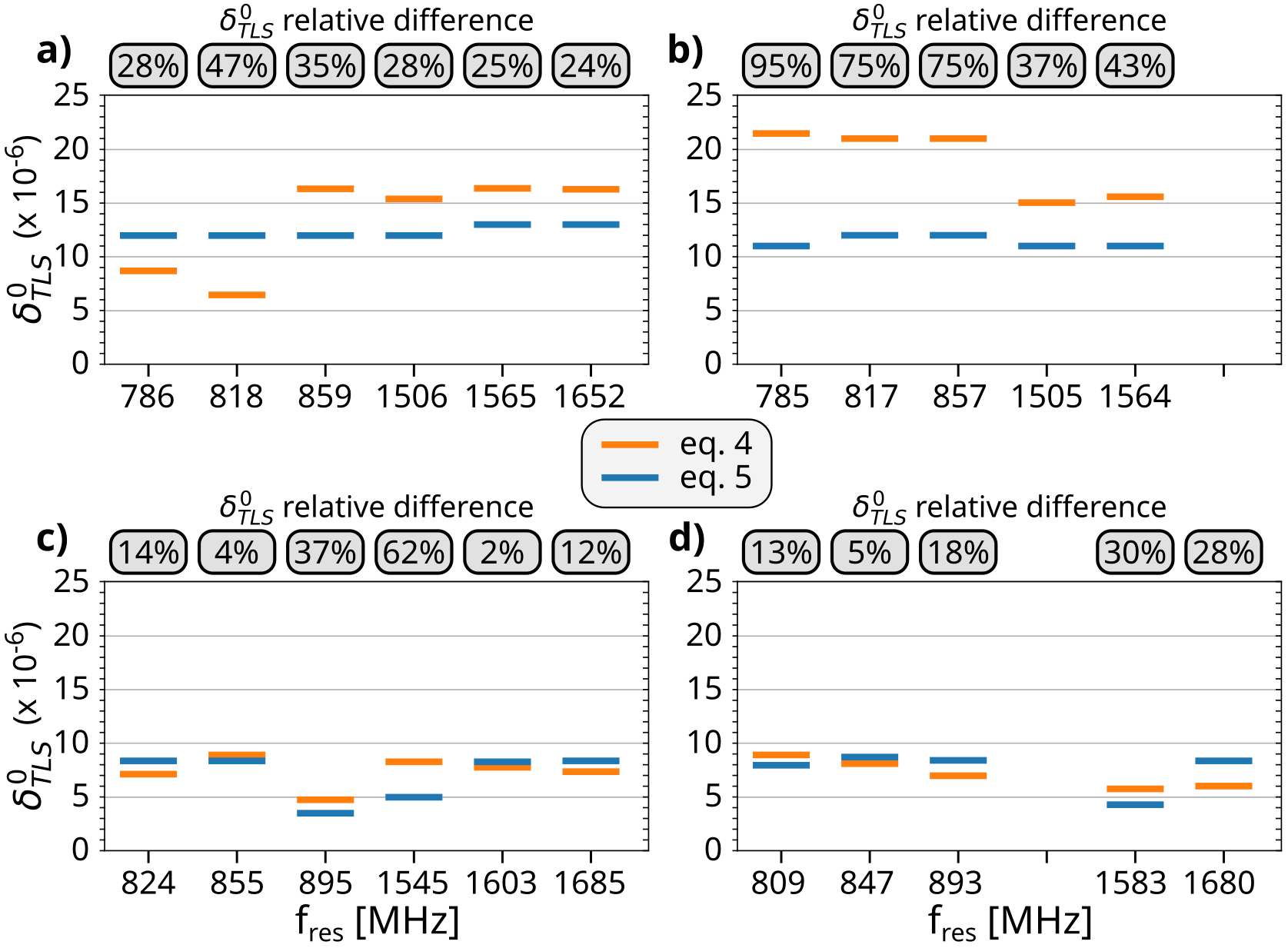}
    \caption{\label{fig:delta0_plots}
      Values of $\delta_{TLS}^0$ obtained from quality-factor data (orange) and frequency-shift data (blue) for a)~Device A(1), b)~Device A(2), c)~Device B(1), d)~Device B(2).  
      The relative difference between the two fitted values of $\delta_{TLS}^0$ is indicated in the grey box above.
      }
\end{figure}

Figure~\ref{fig:loss_fit} shows fits of the measured $f_{res}(T)$ and $Q_i^{-1}(|\vec{E}|)$ to Equations~\ref{eq:gao2} and \ref{eq:loss_tan2}, respectively.  Table~\ref{tab:loss_tan} shows the inferred fit parameters $\delta_{TLS}^0$, $\delta_{other}$, $E_c$, and $\beta$.  Figure~\ref{fig:delta0_plots} summarizes and compares the $\delta_{TLS}^0$ values from all the fits for all the devices.

To calculate the electric field $|\vec{E}|$ from the readout power, we use the following equation, which is derived in Appendix~\ref{app:photon_number}:
\begin{equation}
    \label{eq:Efield2_bis}
    |\vec{E}| = \sqrt{\frac{1}{\epsilon\, A_0\, d}} \sqrt{\frac{P_{read}}{2 \pi f_{res}} \frac{Q_r^2}{Q_c}},
\end{equation}
where $A_0$ is the top plate area, $\epsilon$ is the permitivity of a-Si:H, $d = 800$~nm is the distance separating the plates of the capacitors, and $P_{read}$ is the readout power at the device.  We calculated $A_0 = w \times \ell$ from the dimensions in Table~\ref{tab:reso_dims}.  We assumed $\epsilon/\epsilon_0 = 11.68$, the experimental value found for crystalline silicon, for the a-Si:H relative permitivity because the systematic uncertainty due to any deviation from this value is negligible given the large logarithmic range in $|\vec{E}|$.  Across multiple chips (those presented in this article and others), we measured $f_{res}$ variations smaller than $4\%$.  
Since $f_{res} = 1/(2\pi\sqrt{LC})$ and $C = C_0/2 = \epsilon A_0 / 2d$, the corresponding variation in $\epsilon/d$ would be less than $8\%$.  Inductance variations could also affect $f_{res}$, but simulations indicate that such variations could only occur if the position of the inductor and/or its width varied by more than 2~\textmu m, ruled out by the photolithography accuracy.

\subsection{$\delta_{TLS}^0$}

The best-fit values of $\delta_{TLS}^0$ range from $6.4 \times 10^{-6}$ to $21.5 \times 10^{-6}$ for Devices A(1) and A(2) and from $3.5 \times 10^{-6}$ to $8.9 \times 10^{-6}$ for Devices B(1) and B(2).  The values obtained from frequency shift and quality factor data are within a factor of two of one another.

Among the resonators fabricated using a given recipe (both across resonator frequency and across chips), the frequency-shift values of $\delta_{TLS}^0$ are generally more consistent than the quality-factor values.  This difference arises because there are a number of systematic deviations from expected behavior present only in the quality-factor data and fits while the frequency-shift data seem to follow the model well.  Figure~\ref{fig:loss_fit} shows the $\delta_{TLS}^0$ (low power) plateau is generally visible but the $\delta_{other}$ (high power) plateau is not always, and in some cases, the data are independent of power.  While we can eliminate some of these systematic deviations by excluding the last 5--15 data points, some remain.  We consider three different types of deviations in turn.

In the cases of the Device B(1) resonances at 824~MHz and 895~MHz, we see quality factor independent of readout power over the regime where the model fits the data.  This behavior makes the data insensitive to $\delta_{TLS}^0$, so we consider the $\delta_{TLS}^0$ values for these resonators unreliable and we mark them with an asterisk in Table~\ref{tab:loss_tan}.

The other resonators on Devices B(1) and B(2) show $Q_i^{-1}$ rising at high power (field), presumably due to readout power generation of quasiparticles~\cite{Mazin:2010}.  In the cases of the B(1)~855~MHz and B(2)~847~MHz resonances, the inverse quality factor shows a significant decrease at high power before the rise, while the other Device B(1) and B(2) resonances do not (excluding B(1)~824~MHz and B(1)~895~MHz, already discussed above).  We find the $\delta_{TLS}^0$ values for the B(1)~855~MHz and B(2)~847~MHz resonance quality factor fits are more consistent with their frequency-shift fits, suggesting this difference in behavior causes these resonators' quality factor fits to be more reliable (smaller systematic uncertainty).

The Device A(1) and A(2) resonators display no high-power plateau and in fact show a downward deviation from the expected saturation behavior.  This behavior could be an effect of readout power too, now due to the modification of the quasiparticle distribution function $f(E)$ by readout power~\cite{Guruswamy:2018}.  Independent of the explanation, this behavior makes it impossible to fit for $\delta_{other}$.  

Overall, it is also not known for certain why Devices A(1) and A(2) show primarily a drop in $Q_i^{-1}$ with field while Devices B(1) and B(2) show a rise, though we may speculate that variations in Nb film properties are the cause.

Our conclusion from the above results and discussion is that the frequency-shift data and fits determine $\delta_{TLS}^0$ far more reliably than do quality factor data and fits.  After averaging these values for each recipe, we obtain $\delta_{TLS}^0 \approx 12\,\times 10^{-6}$ for recipe A and $\delta_{TLS}^0 \approx 7\,\times 10^{-6}$ for recipe B.

\subsection{$\delta_{other}$}

The fitted values of $\delta_{other}$ are only valid when a high readout power plateau is visible.  For Devices A(1) and A(2), only the A(1)~786~MHz and A(1)~818~MHz resonances show this plateau, yielding, respectively, values of $7.8 \times 10^{-6}$ and $7.9 \times 10^{-6}$.  For Devices B(1) and B(2), all resonances show the plateau and yield values of $0.2 \times 10^{-6}$ to $28 \times 10^{-6}$.  We see that non-TLS losses vary enormously, even within a single device, and are often non-negligible compared to TLS loss, even at low readout powers.  We do not know for certain the cause of this variation, but we again may propose film property variations.  Certainly, at the low values of $\delta_{TLS}^0$ obtained here, it is clear that non-TLS sources of loss may become important or dominant and thus they deserve future investigation.

\subsection{Evolution of $\delta^0_{TLS}$ and $\delta_{other}$ with time}

We fabricated Devices A(1) and A(2) in October, 2017, and Devices B(1) and B(2) in March, 2020.  We measured Devices A(1) and B(1) twice: first in June, 2018, and September, 2020, respectively, and second in March, 2022.  (We did not remeasure Devices A(2) and B(2).)  Between measurements, we stored the devices in the ambient laboratory atmosphere: air rather than N$_2$ atmosphere; standard air conditioning with generally stable humidity and temperature but no specific additional humidity or temperature control; and, no use of dessicant.

Table~\ref{tab:delta0_evo} shows the $\delta^0_{TLS}$ measurement evolution.  We see a 19\% increase for Device A(1) and a 6\% increase for Device B(1) (both averaged over the resonators on the device).  We can translate these changes to rates of increase per unit time of $0.36 \times 10^{-6}$/month for Device A(1) and $0.33 \times 10^{-6}$/month for Device B(1).  We thus observe that $\delta^0_{TLS}$ increases with time and that the rate of increase is remarkably consistent between the two devices, with an average value of $0.35 \times 10^{-6}$/month.  We may speculate that exposure to laboratory air results in uncontrolled uptake of oxygen, hydrogen, or water by the a-Si:H films that gives rise to an increase in TLS density, but it is also possible that an increase in TLS density arises from evolution of the physical structure of the amorphous films simply due to thermal activation.  We would need to undertake comparisons to a set of devices stored in more controlled environments (vacuum, dry N$_2$ atmosphere, dry air atmosphere, finer temperature control) to narrow down the cause.  These changes are fairly modest, but it is sensible to take simple precautions (\textit{e.g.}, storage with dessicant) to try to prevent such degradation in the future.

\begin{table}[t]
\begin{ruledtabular}
\begin{tabular}{@{}r@{\hskip 3mm}r@{\hskip 3mm}r@{\hskip 6mm}r@{\hskip 3mm}r@{\hskip 3mm}r@{}}
        \multicolumn{3}{l}{Device A(1)} & \multicolumn{3}{l}{Device B(1)} \\ \midrule
        $f_{res}$ & new $\delta^0_{TLS}$ & old $\delta^0_{TLS}$ & $f_{res}$ & new $\delta^0_{TLS}$ & old $\delta^0_{TLS}$ \\ \addlinespace[0.5ex]
        {[}MHz{]} & ($\times 10^{-6}$) & ($\times 10^{-6}$) & {[}MHz{]} & ($\times 10^{-6}$) & ($\times 10^{-6}$) \\ \midrule
        \addlinespace[1ex]
        786 & 12.0 & 10.0 & 824 & 8.3 & 7.5 \\
        818 & 12.0 & 9.9 & 855 & 8.3 & 7.7 \\
        859 & 12.0 & 9.5 & 895 & 3.5 & 4.0 \\
        1506 & 12.0 & 11.0 & 1545 & 5.0 & 4.5 \\
        1565 & 13.0 & N/A & 1603 & 8.1 & 7.5 \\
        1652 & 13.0 & 11.0 & 1685 & 8.3 & 7.6 \\
\end{tabular}
\end{ruledtabular}
        \caption{\label{tab:delta0_evo} Degradation of $\delta_{TLS}^0$ with time as evidenced by $\delta^0_{TLS}$ measurements from frequency-shift data taken at different dates.  For Device A(1), the old and new measurements are separated by 53 months, while, for Device B(1), they are separated by 18 months.
    }
\end{table}

Concerning $\delta_{other}$, the evolution with time seems more random.  In September, 2020, we measured the high power $Q_i^{-1}$ for Device B(1).  
Compared to the $\delta_{other}$ fits obtained from March, 2022, data and presented in Table~\ref{tab:loss_tan}, four resonators show differences smaller than a factor of 2 while two resonators (at 824~MHz and 895~MHz) show increases by factors of 16 and 3, respectively.  We observe that the resonators that show the largest degradations are the ones for which the values of $\delta_{other}$ are largest, absolutely, and large compared to $\delta_{TLS}^0$.

The dramatic differences between both the values and the degradation of $\delta_{other}$ for resonators on the same device suggest that the cause is not general atmospheric or thermal conditions. Local changes in film properties, such as stress, perhaps aggravated by thermal cycling during test, seem a more likely culprit.  Diagnostic measurements sensitive to local film properties --- profilometry or AFM --- may be useful in identifying a specific cause.  It remains to be seen whether these seemingly occasional large absolute values and degradations of $\delta_{other}$ will prove to be a significant practical issue for large arrays of such resonators (\textit{i.e.}, resonator yield).

Jointly, the variations of $\delta_{TLS}^0$ and $\delta_{other}$ with time imply that one must take care to fully re-characterize all resonators during each cooldown: values from previous cooldowns may not be sufficiently accurate.

\subsection{Measurements of $E_c$ and $\beta$}
Recall that $\beta$ is related to the TLS density of states; $\beta = 0.5$ corresponds to a log-uniform density of states.  The values of $\beta$ seen in the literature, varying from 0.15 to 0.35 \cite{Molina:2021, Burnett:2016}, motivate us to also fit for $\beta$ rather than assume the naive value of 0.5.  We require $\beta \in [0.10, 0.60]$ to cover both the naive value and the prior literature values.

The position and slope of the $|\vec{E}|$-dependent region of the $Q_i^{-1}(|\vec{E}|)$ data between the two plateaus (see Figures~\ref{fig:Qi-1_func} and \ref{fig:loss_fit}) determines the values of $E_c$ and $\beta$, respectively, via Equation~\ref{eq:loss_tan2}.  As noted previously, the plateaus are not cleanly visible in all the data, resulting in unreliable values of $E_c$ and $\beta$.  We note these cases with an asterisk in Table~\ref{tab:loss_tan}.  Considering only the reliable fits, we find $0.2 < \beta < 0.6$, with the $\beta = 0.6$ values potentially limited by the allowed range for $\beta$.  The median values of $E_c$ and $\beta$ for Devices B(1) and B(2) for the reliable fits are 7.3~V/m and 0.33, respectively.  The $\beta$ values are thus consistent with the literature.  ($E_c$ values are not available from the literature in general, presumably because, for non-PPC geometries, determining it requires sophisticated modeling of the variation of $|\vec{E}|$ over the resonator at a given $P_{read}$.)  However, we find $E_c$ and $\beta$ to be fairly degenerate, potentially giving rise to the large variation in $\beta$ we see.  Recall that our data only probe values of $h f_{res}/(2 k_B T) \lesssim 0.075$, well away from saturation of the temperature dependence.  Data at lower temperature may thus provide a larger sloped region, improving the determination of $\beta$ and $E_c$.  We thus conclude that our results are consistent with prior data but that we would need lower-$T$ data to constrain $\beta$ more tightly than prior work.

\section{Comparison to Literature Results; Discussion}
\label{sec:comparisons}

To provide context for our results, we review the literature on comparably low-loss, depositable dielectrics.

Comparing these $\delta_{TLS}^0$ results with previously published measurements is not always straightforward, in large part because prior measurements rely almost exclusively on $Q_i^{-1}$ rather than $f_{res}$ data.  Considering resonator design, a non-PPC geometry (\textit{e.g.}, CPW, microstripline) results in a design-specific value of $F$, the TLS fill factor, which is generally calculated and reported to obtain a value of $\delta_{TLS}^0$.  Less frequently, if ever, reported is the $E_c$ value because, as noted above, determining it requires somewhat sophisticated modeling of the variation of $|\vec{E}|$ over the device and how the variation in field saturation with position determines the overall $Q_i^{-1}$.  Considering measurement technique, determining the true value of $\delta_{TLS}^0$ requires either data taken over a sufficiently broad range of $|\vec{E}|$ (usually done) as well as data taken at sufficiently low $T$ such that the temperature dependence saturates (generally but not always possible) or data taken at a wide enough range of $T$ that the temperature-dependence can be fitted for (possible but rarely done).  

We consider in turn other studies of a-Si:H and of other depositable low-loss dielectrics (a-SiC:H and a-Ge).

\subsection{a-Si:H}
\label{sec:comparisons_aSiH}

Among the existing studies of dielectric loss in a-Si:H, the one from O'Connell et al.~\cite{OConnell:2008} is most straightforward to compare because they used PPCs also.  They found $Q_i^{-1} \in [22 - 25] \times 10^{-6}$ for resonance frequencies close to 6~GHz and at 100~mK.  A plot shown in the article also indicates that $\delta_{other} \ll \delta_{TLS}^0$ and that $|\vec{E}| \ll E_c$.  We apply a modest correction for incomplete temperature saturation (from Equation~\ref{eq:loss_tan2}, $Q_i^{-1}(|\vec{E}| \ll E_c) \approx \delta^0_{TLS} \tanh(hf_{res}/(2k_BT))$) to their values of $Q_i^{-1}$ to obtain $\delta^0_{TLS} \in [25 - 28] \times 10^{-6}$.

Mazin et al.~\cite{Mazin:2010} used Al microstrip resonators.  Because they operated below 100~mK and at about 9~GHz, $\tanh(hf_{res}/(2k_BT)) \approx 1$.  Their data show a clear low-power plateau, indicating $\delta_{other} \ll \delta_{TLS}^0$.  We thus extract from their data the value $\delta^0_{TLS} \approx 60 \times 10^{-6}$.  

Bruno et al.~\cite{Bruno:2012} used Nb lumped-element LC resonators.  While measurement details are lacking, it is reasonable to infer that $\delta_{TLS} = 25 \times 10^{-6}$ was obtained at $T = 4.2$~K for $f_{res} = 10.5$~GHz.  The correction for $\tanh(hf_{res}/(2k_BT)) \approx 0.06$ is large, yielding $\delta_{TLS}^0 \approx 420 \times 10^{-6}$.

Molina-Ruiz et al.~\cite{Molina:2021} measured Al CPW resonators on a-Si at 10~mK and 4-7~GHz.  
While this material is not hydrogenated (H is undetectable, with an upper limit of $< 0.1\%$), we include it in discussion of a-Si:H because it can be considered to be on a continuum with a-Si:H.
They also used $Q_i^{-1}$ measurements across a wide range of readout powers to extract $\delta_{TLS}^{0}$.  They tried to correct for the contribution to $\delta_{TLS}^{0}$ of TLS loss of oxide layers at exposed surfaces or at film interfaces by measuring resonators of the same geometry fabricated directly on a bulk crystalline Si substrate, introducing a systematic uncertainty.  The value obtained for their lowest-loss deposition recipe, $\delta_{TLS}^0 = 3.3 \pm 3.5 \times 10^{-4}$, is interestingly low, but it has a large fractional uncertainty for this reason.  We note that this film has low acoustic loss tangent, $21\,\times\,10^{-6}$, though Molina-Ruiz et al.~\cite{Molina:2021} also show that acoustic and RF loss have different microscopic sources, with a thinner film showing lower acoustic loss and higher RF loss, so there is no guarantee that the true RF loss is comparable to the low acoustic loss. 

Buijtendorp et al.~\cite{Buijtendorp:2020, Buijtendorp:2022} and H{\"a}hnle et al.~\cite{Hahnle:2021} provide the most recent results on a-Si:H loss.  While Buijtendorp et al.~\cite{Buijtendorp:2020, Buijtendorp:2022} provide the more extensive study of a-Si:H, their range of readout power is not large enough to identify the low- and high-power plateaus, so $\delta_{TLS}^0$ cannot be accurately determined.  They also make significant corrections for surface oxides that introduce scatter and make it difficult to identify a single number or robust upper limit.  H{\"a}hnle et al.~\cite{Hahnle:2021}, by contrast, provide $Q_i$ data covering the full range needed, yielding $\delta_{TLS}^0 = 3.6\pm 0.5\,\times\,10^{-5}$ (after subtracting $\delta_{other}$; $\tanh(hf_{res}/(2k_BT)) \approx 1$ for $f_{res} = 6$~GHz and $T = 60$~mK).  H{\"a}hnle et al.~\cite{Hahnle:2021} references~\cite{Buijtendorp:2020} for deposition parameters, so we assume the former used the lowest loss recipe from the latter.

It is clear from the above summary that many authors have demonstrated a-Si:H films with $\delta_{TLS}^0 < 10^{-4}$.  It is of obvious interest to identify under what conditions these results can be obtained and what causes the remaining variability.  We summarize in Table~\ref{tab:fab_recipes_compa} deposition recipes and measured $\delta_{TLS}^0$ for our devices and the above literature results.  It is clear that $\delta_{TLS}^0 < 10^{-4}$ seems achievable with a variety of PECVD techniques and recipes and that these various recipes exhibit less than an order of magnitude variation in $\delta_{TLS}^0$.  However, no patterns connecting $\delta_{TLS}^0$ to PECVD deposition technique or parameters readily emerge.  

Due to the large uncertainty on $\delta_{TLS}^0$ for e-beam evaporated a-Si films, it is not yet clear whether such films offer comparably low $\delta_{TLS}^0$.   Molina-Ruiz et al.~\cite{Molina:2021} show that a lower substrate temperature during deposition dramatically increases both RF and acoustic loss, while we do not observe a similar dependence for PECVD films, so it does seem that PECVD is a more robust process for achieving $\delta_{TLS}^0 < 10^{-4}$ for a-Si:H.  (We note that PECVD films of SiO$_2$ and SiN$_x$ exhibit significantly higher loss tangents, with the previously noted exception of the high-stress film in~\cite{Paik:2010}, so PECVD alone is not sufficient.)

This comparison with the previously published measurements shows that the average value of a-Si:H $\delta_{TLS}^0$ for our recipe B is about 4 times lower than that measured by O'Connell et al.~\cite{OConnell:2008}, 8 times lower than that measured by Mazin et al.~\cite{Mazin:2010}, of order a decade lower than that measured by Molina-Ruiz et al.~\cite{Molina:2021}, and 5 times lower than that measured by Hahnle et al.~\cite{Hahnle:2021}.  
\textit{We thus believe we have obtained the lowest-loss a-Si:H films in the literature.  Moreover, the modestly poorer loss obtained from recipe A implies our $\delta_{TLS}^0$ is fairly robust against changes in deposition machine and recipe.}

\begin{table*}[t]
\begin{ruledtabular}
\begin{tabular}{lrrrrrrrrr}
        Ref / Facility & Process & Gas ratio & Temperature & Gas pressure & Gas flow & Deposition time & Thickness & $\delta_{TLS}^0$ \\ 
         &  & (SiH\textsubscript{4}/Ar) & {[}\textdegree C{]} & {[}mTorr{]} & {[}sccm{]} &   & {[}nm{]}  &  ($\times 10^{-6}$) \\ \midrule
        Caltech KNI & PECVD & 5\% / 95\% & 350 & 800 & 250 & 27'11" & 800 & $\approx 12$ \\
        JPL MDL & ICP-PECVD & 100\% / 0\% & 350 & 10 & 30 & 26'06" & 800 & $\approx 7$ \\ 
        \addlinespace[1ex]
        O’Connell et al.~\cite{OConnell:2008} & HD PECVD & 66\% / 33\% & 100 & 2 & 45 & 3'15" & unknown & $\approx 26$ \\
        Mazin et al.~\cite{Mazin:2010} & HD PECVD & 66\% / 33\% & 100 & 2 & 45 & 3'15" & 200 & $\approx 60$\\
        Bruno et al.~\cite{Bruno:2012} & PECVD    & 100\% / 0\% & 250 & 200 & unknown & unknown & unknown & 420 \\
        Molina-Ruiz et al.\,(no H)~\cite{Molina:2021} & e-beam evap & N/A & 50 / 225 / 450 & $<5 \times 10^{-6}$ & N/A & 19'-106' & 59-317 & $\ge 330 \pm 350$ \\
        H{\"a}hnle et al.~\cite{Hahnle:2021} & PECVD & 5\% / 95\% & 350 & 1000 & 500 & 7'0" & 250 $\pm$ 15 & 36$\pm$5 \\
        \end{tabular}
\end{ruledtabular}
        \caption{\label{tab:fab_recipes_compa} Summary of a-Si:H deposition recipes and measured $\delta_{TLS}^0$ for our devices and previously published results.  a-Si:H films measured by Mazin et al.\ and O'Connell et al.\ were both fabricated at University of California Santa Barbara (UCSB) using a recipe identical or very close to the one detailed by E.~Lucero in his PhD thesis~\cite{Lucero:2012}, Appendix B.4.6 (private communication from B.~Mazin).  HD PECVD stands for High-Density Plasma-Enhanced Chemical Vapor Deposition.  
        While most of the above publications use PECVD to deposit a-Si:H, Molina-Ruiz et al.\ use e-beam evaporation instead to deposit a-Si and thus gas ratio and gas flow are not applicable.  O'Connell et al.\ do not provide precise film thicknesses but indicate their films are hundreds of nm thick, comparable to the other films presented in this table.  
        As noted in the text, Buijtendorp et al.~\cite{Buijtendorp:2020, Buijtendorp:2022} and H{\"a}hnle et al.~\cite{Hahnle:2021} use the same deposition parameters, but the former do not provide precise $\delta_{TLS}^0$ values, while the latter does not provide deposition details.  We assume H{\"a}hnle et al.~\cite{Hahnle:2021} used the lowest loss recipe from Buijtendorp et al.~\cite{Buijtendorp:2020}.
        }
\end{table*}

\subsection{Other Depositable Low-Loss Dielectrics: a-SiC:H and a-Ge}

While our focus in this paper has been on a-Si:H because, out of the widely studied depositable dielectrics AlO\textsubscript{x}, SiO\textsubscript{2}, SiO\textsubscript{x}, SiN\textsubscript{x}, and a-Si:H, it generally gives the lowest loss tangents, there are other depositable dielectrics with comparably low loss and thus of potentially comparable utility.  

Buijtendorp et al.~\cite{Buijtendorp:2022} studies hydrogenated amorphous silicon carbide, a-SiC:H, deposited using PECVD.  They measure $\delta_{TLS}^0 = 30\pm0.4 \times 10^{-6}$ at $f_{res} = 7.4$~GHz and $T = 60$~mK, requiring no $\tanh(hf_{res}/(2k_BT))$ correction.  These results are comparable to the same authors' results for a-Si:H in H{\"a}hnle et al.~\cite{Hahnle:2021}, noted above.  

Kopas et al.~\cite{Kopas:2021} have demonstrated similar low loss tangents for a-Ge.  They use Nb CPW resonators deposited on a 1~\textmu m thick a-Ge film deposited by thermal evaporation.  The temperature (40~mK) and resonance frequency (6.3~GHz and 7.3~GHz) of their measurements ensure that $\tanh(hf_{res}/(2k_BT)) \approx 1$.  Their range of readout power reaches the low-power plateau, so their low-power loss tangent measurement $Q_i^{-1} = [11 - 13] \times 10^{-6}$ corresponds to $\delta_{TLS}^0 + \delta_{other}$.  Their data do not clearly reach the high-power plateau that would yield $\delta_{other}$, so the measurements are an upper limit on $\delta_{TLS}^0$.  Because the measured loss tangent is so low, they subtract contributions from TLS at various interfaces and in the substrate based on calculated fill factors and literature values~\cite{Woods:2019}: $5.8 \times 10^{-6}$, $1.1 \times 10^{-6}$, $0.8 \times 10^{-6}$, and $2.6\,\times\,10^{-7}$ for the metal-substrate, substrate-air, and metal-air interfaces and silicon substrate loss, respectively.  They obtain $\delta_{TLS}^0 + \delta_{other} = [4.7 - 4.9] \times 10^{-6}$ for a-Ge alone, which we interpret as an upper limit on $\delta_{TLS}^0$.  However, no uncertainties on these large subtractions are provided, and it is not clear that these subtractions are valid, as~\cite{Woods:2019} employed TiN resonators on crystalline Si while these authors consider Nb resonators on an a-Ge film on crystalline Si.  Thus, $\delta_{TLS}^0 < 11 \times 10^{-6}$ is the more robust result.

\section{Conclusion}
\label{sec:conclusion}

We have reported low-field, low-temperature TLS loss ($\delta_{TLS}^0$) values of 7$\,\times 10^{-6}$ and 12$\,\times 10^{-6}$ for two similar but far from identical deposition recipes, obtained in two different CVD deposition systems at two different sites.  These values are robust and subject to minimal modeling uncertainty thanks to the use of PPCs, multiple resonators spanning a factor of 2 in frequency, and frequency-shift data.  We have also determined $\delta_{TLS}^0$ from quality-factor data and found those values to be broadly consistent with the frequency-shift results, but we have explained why quality-factor data is more difficult to model and thus the resulting values subject to greater systematic uncertainty.  Our lowest frequency-shift values of TLS loss are a factor of 4 lower than the best previously published, robust results for a-Si:H and are comparable to results for a-Ge \cite{Kopas:2021}.  
They also depend only modestly (factor of $\approx$1.5) on deposition machine and details of deposition technique.  Comparison to the literature reveals no clear reason for the very low observed $\delta_{TLS}^0$ values, though it does suggest that CVD is a robust process for achieving $\delta_{TLS}^0 < 10^{-4}$ with a-Si:H.  A variety of fields employing superconducting devices with dielectric film capacitors may benefit from this development.  

We have also characterized, with varying levels of robustness, the other parameters $\delta_{other}$, $E_c$, and $\beta$ that impact the quality factor of resonators using these films.  We observed degradation of both $\delta_{TLS}^0$ and $\delta_{other}$ with time when no special storage measures are taken, motivating more precautions in the future.

Lastly, we have demonstrated that the technique of using frequency-shift data with PPC resonators is a highly robust way to study TLS loss and that the use of quality-factor data requires careful attention to the temperature and field values at which the data are taken.  In the latter case, to infer $\delta_{other}$, $E_c$, and $\beta$ --- the latter two of which inform us about the microscopic nature of the TLS --- a large dynamic range in $|\vec{E}|$ must be probed, covering both the low and higher power plateaus.  Other effects (quasiparticle loss and/or heating) may make such analyses challenging.

Future publications in preparation will study: the dependence of the above TLS and non-TLS parameters on deposition parameters in the machines used for this work; and, correlations of TLS loss and TLS noise for these recipes.  Future work underway will include: comparisons of these parameters for resonators fabricated using other machines at other sites but with efforts made to mimic deposition recipes; correlation of TLS loss, noise, and room-temperature film-quality diagnostics; and, correlation of RF ($\sim 1$~GHz) and mm-wave ($\sim 100$--400~GHz) loss.

Future work will study: the dependence of the above TLS and non-TLS parameters on deposition parameters in the machines used for this work; comparisons of these parameters for resonators fabricated using other machines at other sites but with efforts made to mimic deposition recipes; correlation of RF ($\sim 1$~GHz) and mm-wave/THz ($\sim 100$--400~GHz) loss; correlation of RF loss and RF noise; and, correlation of RF loss and noise with film properties measured at 300~K (O/IR refractive index and absorption, density, H content and elemental composition including binding state, bond types and angles, film strain, short-range and medium-range order, dangling bond density, and perhaps others) and with cryogenic acoustic loss.  These efforts will have both practical and fundamental consequences.  At a practical level, the study of deposition conditions will test the robustness of our recipes and enable reproduction by other investigators.  The correlation with 300~K diagnostics will offer a means to understand the fundamental reasons for the low RF loss from underlying material properties as well as a practical way to monitor and tune deposition processes to maintain the low loss (and perhaps further improve it).  Noise measurements will test whether, at a practical level, low loss a-Si:H films also will enable low-noise PPC KIDs while also offering another empirical constraint on the fundamental tunneling model explanation of a-Si:H behavior.  Mm-wave/THz measurements will test the material's practical utility in that frequency regime as well as probe the electric-dipole-couple TLS density of states at 100$\times$ higher energies.  Cryogenic acoustic loss measurements will test whether the conclusion of Molina-Ruiz~et~al.~\cite{Molina:2021}, that two types of TLS are present in a-Si, is valid at the 25--50$\times$ lower RF loss levels observed here; perhaps, there is one species of electric-dipole-coupled TLS that do not interact acoustically while the acoustically active species is also electric-dipole-coupled but more weakly than the former.

\begin{acknowledgments}
This work has been supported by the JPL Research and Technology Development Fund, the National Aeronautics and Space Administration under awards 80NSSC18K0385 and 80NSSC22K1556, and the Department of Energy Office of High-Energy Physics Advanced Detector Research program under award DOE-SC0018126.  A. B. and F. D. carried out research/fabrication at the Jet Propulsion Laboratory, operated by the California Institute of Technology under a contract with the National Aeronautics and Space Administration (80NM0018D0004).  The authors would like to thank J.~Gao for his very helpful comments, B.~Mazin for efforts to re-make devices using the O'Connell et al.~\cite{OConnell:2008} recipes, and M.~Hollister for design and construction of the cryostat used for this work and early resonator measurements.

\end{acknowledgments}

\bibliography{biblio}

\appendix

\section{Literature on a-Si and a-Si:H}
\label{app:review}

As noted in \S\ref{sec:intro}, there is significant, potentially relevant literature on a-Si and a-Si:H in three different contexts: as a photovoltaic, as an example of sub-electronic-bandgap optical absorption, with impact on its use as a mirror coating in laser-interferometric gravitational wave detectors (GWDs); and, as an exemplar of universal STM behavior.  The common thread in the literature of relevance to our work is the impact of dangling bonds in the nominally four-fold coordinated a-Si material, quantifiable via electron paramagnetic resonance measurements.  While the photovoltaic studies (see~\cite{Crandall:1995} for a review) devote significant attention to the importance of dangling bonds to photovoltaic efficiency and the role of hydrogenation in both improving and degrading that efficiency through its effect on the dangling bond density, the electromagnetic behavior studied is at optical frequencies, where the processes involved are energetically disparate from those relevant here (eV vs.\ $\mu$eV--meV).  The same holds for GWDs: the concern is with optical absorption in the 1--2~$\mu$m (150--300~THz; 0.5-1~eV) regime.  While there are studies correlating this absorption with dangling bond density at high absorption (extinction coefficient $\gtrsim 10^{-3}$), the absorption at the lower levels of practical interest for GWDs appears to be due to sub-electronic-bandgap states, not dangling bonds~\cite{Birney:2018}.  

The third context, the study of a-Si and a-Si:H as an example of the STM is, in contrast, quite qualitatively relevant: the modern introduction of a-Si/a-Si:H for GHz superconducting device applications by O'Connell et al.~\cite{OConnell:2008} identified a-Si/a-Si:H as the most promising easily deposited material from the extensive database of thermal conductivity and acoustic attenuation data in amorphous materials presented in Pohl et al.~\cite{Pohl:2002}, whose goal was to assess the universality of the low-energy excitations on which the STM is based.  However, even there, it is now clear the connection to that extensive prior literature is tangential: Molina-Ruiz et al.~\cite{Molina:2021} demonstrated that, at least in some regimes of TLS density, there appear to be two types of TLS: some that couple strongly to elastic strain (phonons) and some that couple strongly via electric-dipole interaction to electromagnetic fields.  We are of course interested in the latter, while the literature on a-Si/a-Si:H prior to Molina-Ruiz et al.~\cite{Molina:2021} was primarily concerned with the former.  At the moment, this two-TLS model suggests the extensive prior literature is not quantitatively relevant for the work here, though we offer a caveat in \S\ref{sec:conclusion}.

\section{Electric field and photon number calculations}
\label{app:photon_number}

\begin{figure*}[t]
  \includegraphics[width=17.5cm]{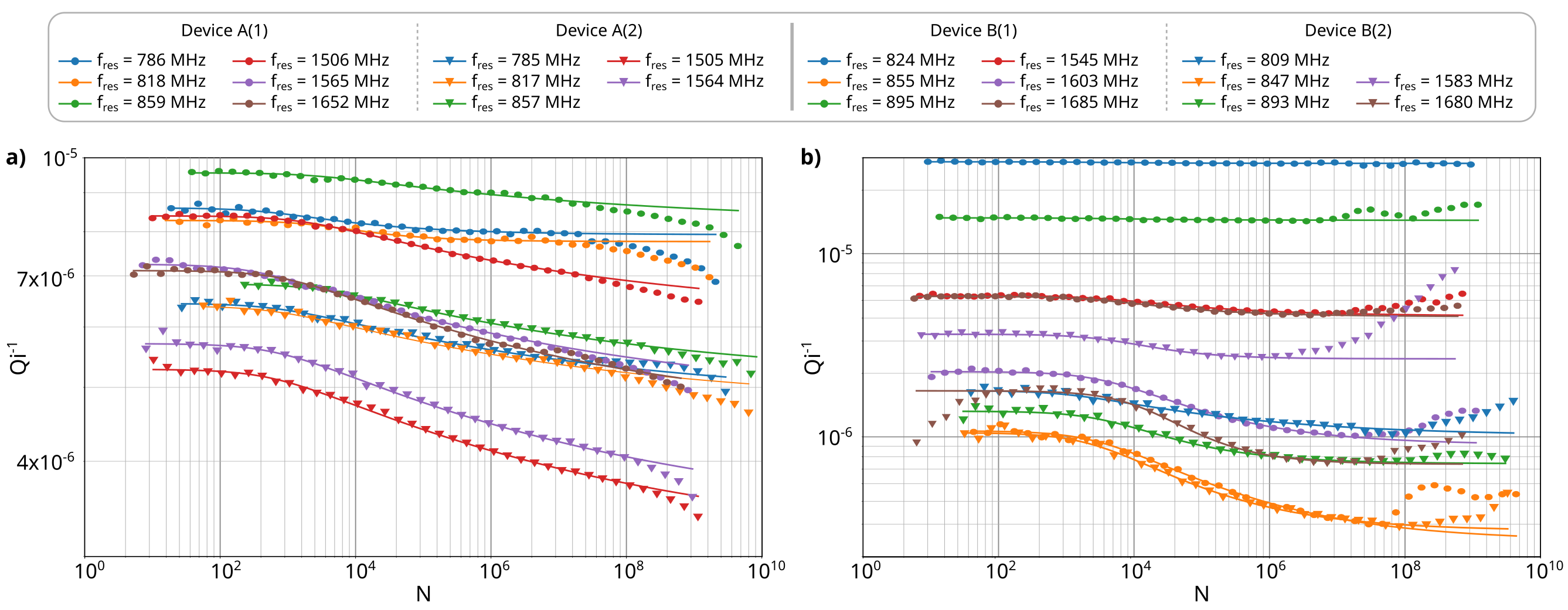}
    \caption{\label{fig:loss_fit_vs_Nphotons} a),~b)~Data and best-fit TLS model (Equation~\ref{eq:loss_tan2}) for $Q_i^{-1}$ data as a function of photon number.  The data shown on this plot 
    are the same as in Figure~\ref{fig:loss_fit} c)~\&~d) but $|\vec{E}|$ has been converted to photon number $N$ using Equation~\ref{eq:photon_N}.}
\end{figure*}

The total energy stored in a resonator at a frequency $f$ can be expressed as \cite{Swenson:2013, Siegel:2015, Zmuidzinas:2012}: 
\begin{equation}
    \label{eq:Energy_global}
    W = \frac{1}{1+4Q_r^2x^2}\frac{P_{read}}{\pi f_{res}} \frac{Q_r^2}{Q_c},
\end{equation}
with $x = (f-f_{res})/f_{res}$, and $P_{read}$ the readout power at the device.
At $f = f_{res}$ ($x = 0$), this equation simplifies to:
\begin{equation}
    \label{eq:Energy_fres}
    W = \frac{P_{read}}{\pi f_{res}} \frac{Q_r^2}{Q_c}.
\end{equation}
The total time-averaged stored energy $W$ is composed of time-averaged magnetic and electric energies, respectively, which are identical and half of the total energy: $W_e = W_m = W/2$.  Only the electric energy $W_e$ should be considered for calculating the rms electric field magnitude (which we denote, for brevity, by $|\vec{E}|$) between the parallel plates of the capacitor.  The generic formula for a single parallel-plate capacitor is:
\begin{equation}
    \label{eq:Efield}
    |\vec{E}| = \sqrt{\frac{2 W_e}{\epsilon A d}},
\end{equation}
where $\epsilon$ is the permitivity of the dielectric between the capacitor plates, $A$ is the area of each capacitor plate, and $d$ is the separation of the plates (the thickness of the dielectric film). 
However, in our specific case, because we have two identical capacitors in series, the electric energy stored in each of them is $W_e/2$, so we can rewrite Equation~\ref{eq:Efield} as:
\begin{equation}
    \label{eq:Efield2}
    |\vec{E}| = \frac{\sqrt{\langle V^2 \rangle}}{d} = \frac{1}{d} \sqrt{\frac{W_e}{C_0}} = \sqrt{\frac{1}{\epsilon A_0 d}} \sqrt{\frac{P_{read}}{2 \pi f_{res}} \frac{Q_r^2}{Q_c}},
\end{equation}
where $A_0$ is the top plate area and $C_0$ is the capacitance of each of the two capacitors in series.  We use the above formula to convert the applied $P_{read}$ to the $|\vec{E}|$ values used for our plots and fits.

The energy stored in a resonator can also be converted into a number of stored photons.  In the literature, photon number is frequently reported in place of readout power.  The relation between the two is (also used by Molina-Ruiz et al.~\cite{Molina:2021}):
\begin{equation}
    \label{eq:photon_N}
    N = \frac{W}{h f_{res}} = \frac{P_{read}}{\pi h f_{res}^2} \frac{Q_r^2}{Q_c},
\end{equation}
where $W = W_e + W_m$ is the total electric and magnetic energy stored in the resonator and $h$ is the Planck constant.  
We use it to plot $Q_i^{-1}$ data as a function of photon number in Figure~\ref{fig:loss_fit_vs_Nphotons} to enable comparisons on that basis.  
Photon number has no special relevance in this work, as we believe $E_c$ is a more fundamental microscopic parameter.

\end{document}